# Material Properties at Low Temperature

*P. Duthil*[1]

Institut de Physique Nucléaire d'Orsay, IN2P3-CNRS/Université de Paris Sud, Orsay, France

**Abstract**

From ambient down to cryogenic temperatures, the behaviour of materials changes greatly. Mechanisms leading to variations in electrical, thermal, mechanical, and magnetic properties in pure metals, alloys, and insulators are briefly introduced from a general engineering standpoint. Data sets are provided for materials commonly used in cryogenic systems for design purposes.

## 1 Introduction

The properties of materials are used in specific equations to predict the behaviour of a system. They depend on chemical composition, crystalline architecture, atomic and charge interactions, heat or mechanical treatments, etc., and are affected by temperature changes. Thus, understanding the mechanisms that contribute to relevant properties is important when considering changes in materials with temperature. For details about transport phenomena properties in matter (e.g. thermal or electrical properties), one can refer to any textbook on solid-state physics (see, for example, Refs. [1, 2]). Magnetism or superconducting states also affect material properties. The reader is encouraged to refer to dedicated lectures (which are too specific to be presented here) at the CERN Accelerator School. References on data concerning materials used in cryogenics are given in Ref. [3].

## 2 Thermal properties

The thermal (and electrical) properties of any material are related to the vibrations of its atoms around their equilibrium positions (in a lattice crystal). The amplitude of these vibrations depends on temperature and diminishes as the temperature decreases. Note that these vibrations may propagate within the material at the speed of sound, and are studied as plane waves, with which phonons are associated. Thermal properties also depend on the movements of negative charges (electrons) and positive charges (vacancies) if the material is a conductor.

### 2.1 Heat capacity

The heat capacity $C$ is defined as the quantity of energy (heat) that must be introduced into some mass of material to increase its temperature $T$ by 1 K. Reversibly, it is the quantity of energy extracted from this mass of material to decrease its temperature by 1 K. The unit of this extensive quantity is $J \cdot K^{-1}$. The intensive property is given by the specific heat $c$, which is the heat capacity or thermal capacity per unit of mass ($J \cdot kg^{-1} \cdot K^{-1}$) or the molar heat capacity ($J \cdot mol^{-1} \cdot K^{-1}$). The heat capacity of a material is thus its ability to store or release heat energy. It is an important property in cool-down or warm-up processes used in the estimation of the energy (and cost) involved, and in the assessment of the transient states of the heat transfer as it relates to the thermal diffusivity (see Section 2.2.5). As the temperature tends to zero, so does the heat capacity.

---
[1] duthil@ipno.in2p3.fr

Heat can be supplied at constant volume $V$ or constant pressure $p$, defining two heat capacities such that:

$$C_V = T\left(\frac{\partial S}{\partial T}\right)_V = \left(\frac{\partial U}{\partial T}\right)_V, \quad C_p = T\left(\frac{\partial S}{\partial T}\right)_p = \left(\frac{\partial H}{\partial T}\right)_p, \quad (1)$$

where $S$ is the entropy, $U$ is the internal energy, and $H$ is the enthalpy (all having units of joules, J). For a system comprising the association of several homogenous materials $i$ each having a mass $m_i$:

$$C_V = \sum_i m_i c_{V_i} \text{ and } C_p = \sum_i m_i c_{p_i}. \quad (2)$$

For solids at low temperatures, the difference $C_p - C_V$ is negligible (less than 1%), and thus no distinction will be made in the following between $C_V$ and $C_p$.

### 2.1.1 *Crystal lattice contribution*

Considering the sum of the energies of the phonon modes, the Debye model enables the expression of the total energy of the crystal lattice as a function of the Debye temperature $\theta_D$:

$$U = 9Nk_B \left(\frac{T}{\theta_D}\right)^3 \int_0^{\theta_D/T} \frac{x^3}{e^x - 1} dx = 3R\, D_3\left(\frac{\theta_D}{T}\right), \quad (3)$$

with

$$\theta_D = \frac{\hbar v_{ph}}{k_B}\left(\frac{3N}{4\pi V}\right)^{1/3}, \quad D_3 = \frac{3}{x^3}\int_0^x \frac{t^3}{e^t - 1} dt, \quad (4)$$

where $k_B = 1.38 \cdot 10^{-23}$ J·K$^{-1}$ is the Boltzmann constant, $\hbar = 1.055 \cdot 10^{-34}$ J·s is the reduced Planck constant, $v_{ph}$ is the speed of sound in the material, $N/V$ is the number of atoms per volume of material, and $D_3$ is called the 'third Debye function'. The Debye temperature of some materials is given in Table A.1 of Appendix A.

For $n = 1$ mole of material, noting that $R = k_B N_A$ and that $N = nN_A$, the derivation of Eq. (3) with respect to the temperature gives the crystal lattice contribution to the molar heat capacity $c_{ph}$ (the subscript 'ph' refers to phonons):

$$\frac{c_{ph}}{3R} = 3\left(\frac{T}{\theta_D}\right)^3 \int_0^{\theta_D/T} \frac{x^4 e^x}{(e^x - 1)^2} dx. \quad (5)$$

It follows that, for any material, the ratio $c_{ph}/(3R)$ can be plotted as a unique function of $T/\theta_D$, as shown in Fig. 1. As $T$ increases, $c_{ph}/(3R)$ tends to 1 (see the black solid curve of Fig. 1) and thus $c_{ph} \to 3R$. For $T \ll \theta_D$, the limit of $c_{ph}$ can be found to be $(T/\theta_D)^3 12\pi^4/5$, showing a cubic dependence of the lattice heat capacity with respect to the temperature (see the grey dashed line of Fig. 1).

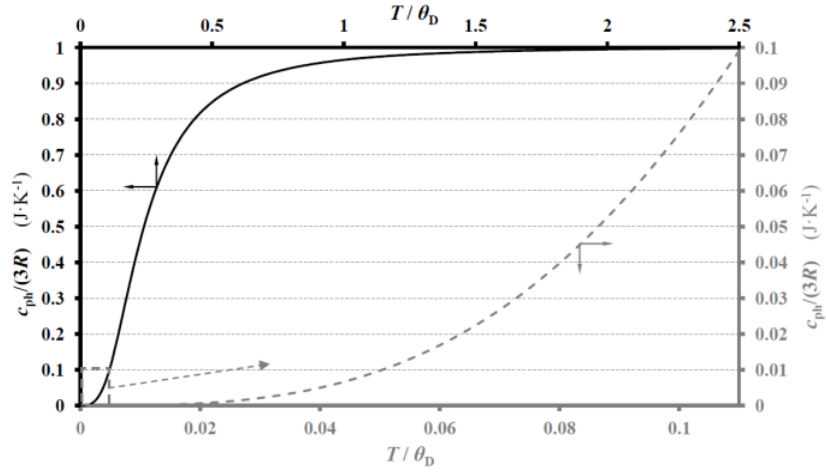

**Fig. 1:** $c_{ph}/(3R)$ as a function of $T/\theta_D$; solid black curve and black axis for $T/\theta_D \in [0; 2.5]$; dashed grey curve and grey axis for $c_{ph}/(3R) \in [0; 0.01]$.

### 2.1.2 Contribution of free electrons

Thermal excitation induced by heating concerns only a fraction of the electrons for which the energy is within $k_B T$ of the Fermi energy, $E_F = (\hbar/2m_e)(3\pi^2 N/V)^{2/3}$, where $m_e = 9.109 \cdot 10^{-31}$ kg is the mass of an electron. The total energy of the electrons is $U \approx (N \cdot T/T_F)k_B T$, where $T_F = E_F/k_B$ is called the 'Fermi temperature'. The electronic heat capacity given by the free-electron-gas model thus evolves linearly with respect to the temperature:

$$C_e = \partial U/\partial T \approx (T/T_F) N k_B \ (\text{unit: J·K}^{-1}). \tag{6}$$

At temperatures much lower than the Fermi temperature, the molar heat capacity (of free electrons) can be expressed by $C_e = \frac{1}{2} \cdot \pi^2 N k_B T / T_F$.

### 2.1.3 Heat capacity of materials

The heat capacities of different materials (metals, thermal insulators, and gases for comparison) are plotted in Fig. 2. Reference [4] provides the heat capacities of some materials as functions of the temperature. The tendencies of the curves are explained in the following sections.

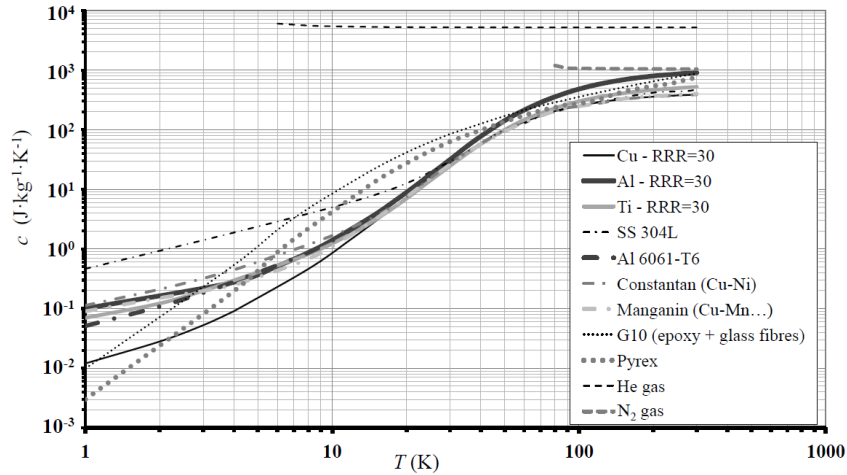

**Fig. 2:** Specific heat $c$ (J·kg$^{-1}$·K$^{-1}$) of some materials versus the temperature $T$ compiled from Ref. [5]. Metals are plotted with solid lines, metallic alloys with dot-dashed lines, and thermal insulators with dotted lines. For comparison, the specific heats of two gases commonly used in cryogenics are plotted with dashed lines.

*2.1.3.1 Heat capacity of metals*

The heat capacity of metals can be seen as the sum of the contributions from the crystal lattice and the free electrons. Thus, at low temperature ($T < \theta_D/10$ and $T \ll T_F$), $C = C_e + C_{ph} = \gamma T + AT^3$, where $A$ and $\gamma$ are constants related to the material. Experimentally, $C/T$ is plotted versus $T^2$; this yields a linear line with slope $A$ and intercept $\gamma$. Some values of $\gamma$ are given in Table A.2. For higher temperatures ($T > 2\theta_D$), $c_{ph} \sim 3R$, and $C$ increases linearly as the temperature rises.

Below 10 K ($c_{ph} \ll 1$), the linear electronic term dominates, and thus the heat capacity is proportional to the temperature.

*2.1.3.2 Heat capacity of thermal insulators*

For crystallized thermal insulators, the contribution of the phonons is predominant. Therefore, $c_{ph} \sim 3R$ for $T > 2\theta_D$ and $c_{ph} = f(T^3)$ for $T < \theta_D/10$.

*2.1.3.3 Heat capacity of superconductors*

For superconductors at temperatures below their critical temperature $T_C$, electrons are linked in Cooper pairs and do not contribute to energy transport. The heat capacity can then be written using BCS theory as [6]:

$$c_s = \gamma T_C a \cdot e^{\left(-bT_C/T\right)} \text{ for } T < T_C, \tag{7}$$

where $\gamma$ is the constant coefficient of the electronic contribution measured at $T > T_C$, and $a$ and $b$ are other constants that depend on the material.

*2.1.4 Sensible heat*

During a thermodynamic process at constant pressure between two temperatures $T_1$ and $T_2$, one can write the specific enthalpy change (unit J·K$^{-1}$) as follows:

$$\Delta h = \int_{T_1}^{T_2} c_p \, dT.$$

Thus, the energy involved in the process is $E = \Delta H = m \cdot \Delta h$; it is called 'sensible heat', referring to a 'sensible' change of temperature of the material. The parameter $\Delta h$ is seen as a heat stock per unit of mass. The enthalpy $h = \int_0^T c_p dT$ is plotted versus temperature for different materials in Fig. 3.

At low temperature, it can be seen that the enthalpies of solid materials, such as G10 (which is a thermal insulator made of epoxy resin and glass fibres) and stainless steel AISI 304L (EN 10088-1&2 / 1.4306) for example, are much smaller than those of gases and especially than that of helium. A consequence is that cooling down materials by removing heat with cold helium vapours (which then warm up) is more efficient than by vaporizing liquid helium (the vaporization latent heat of helium-4 at a pressure of 1 bar is about 20.9 J·g$^{-1}$) as smaller quantity of cold helium is thus involved in that process.

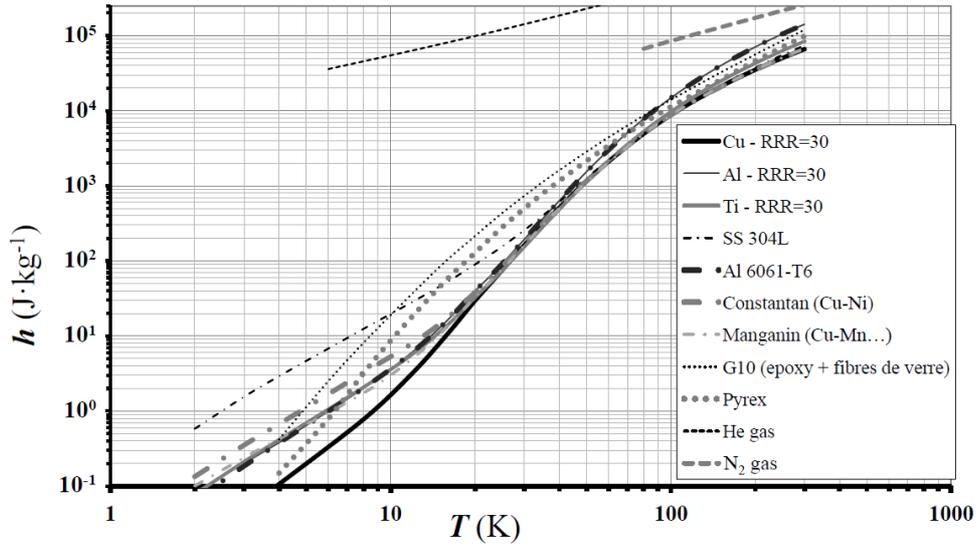

**Fig. 3:** Enthalpy $h = \int_0^T c_p dT$ (J·kg$^{-1}$) of various materials versus temperature $T$. Data were compiled from Ref. [5]. Metals are plotted with solid lines, metallic alloys with dot-dashed lines, and thermal insulators with dotted lines. For comparison, the specific sensible heats of two gases are plotted with dashed lines.

## 2.2 Thermal conductivity

The thermal conductivity $k$ (unit W·m$^{-1}$·K$^{-1}$) relates to the facility with which heat can diffuse into a material. Fourier's law yields the quantity of heat diffusing through a unit surface during a unit of time within a material subjected to a temperature gradient:

$$\vec{\dot{q}} = -k\vec{\nabla T} \quad (\text{J·s}^{-1}\text{·m}^{-2} \equiv \text{W·m}^{-2}). \tag{8}$$

As an example, we can consider the linear (one-dimensional) support of Fig. 4, with length $L$, cross-section $A$, and with its ends at temperatures $T_H$ and $T_C$ (subscripts H and C refer to hot and cold, respectively). Applying Fourier's equation, writing $\dot{Q}_{HC} = A\dot{q}_{HC}$, which is a constant along the linear support (conservation of energy), we obtain

$$\int_0^L \frac{\dot{Q}_{HC}}{A} dx = \dot{Q}_{HC} \frac{L}{A} = -\int_{T_H}^{T_C} k(T) dT \quad . \tag{9}$$

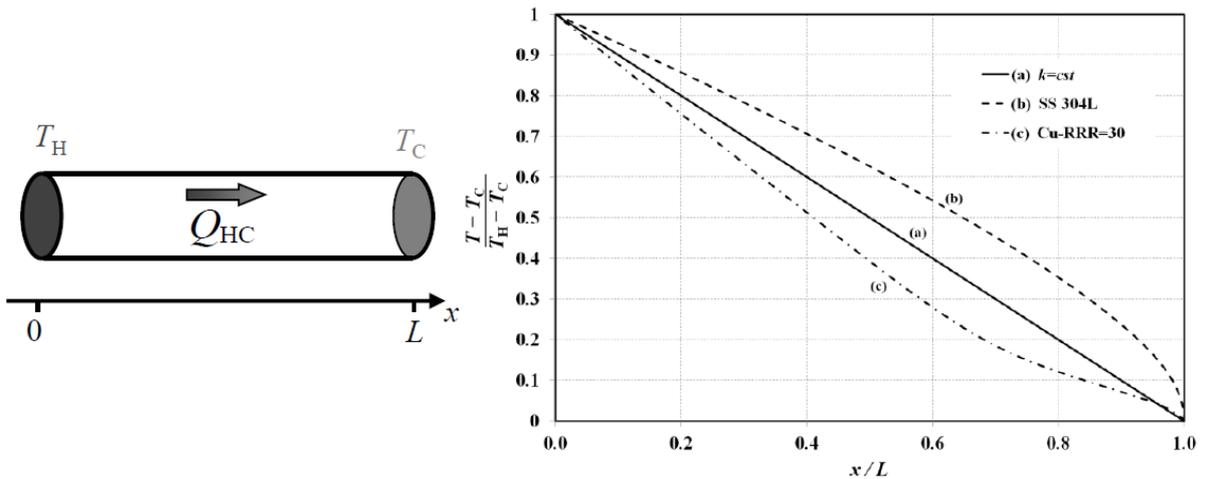

**Fig. 4:** Conduction within a linear support: considered problem (left) and possible temperature solutions depending on the thermal conductivity $k$ (right).

Thus, if $k$ is constant, the quantity of heat per time unit diffusing within the support is given by $\dot{Q}_{HC} = kA(T_H - T_C)/L$ (unit: W), which is a positive quantity as the heat diffuses along the positive $x$-direction from the higher to the lower temperature. The resulting temperature profile $T(x)$ is linear. However, the thermal conductivity depends on temperature, especially in the cryogenic domain, leading to non-linear profiles (see Fig. 4). Simply put, heat is transported in solids by electrons and phonons, and the thermal conductivity is seen as the sum of both contributions:

$$k = k_e + k_{ph}.$$

### 2.2.1 Lattice contribution to the thermal conductivity

Considering the energy change $c\Delta T$ of a particle travelling over a mean free path $\ell_{ph}$ between two temperatures $T$ and $T + \Delta T$, the phonon contribution to the thermal conductivity $k_{ph}$ is given by

$$k_{ph} = \frac{1}{3} c_{ph} v_{ph} \ell_{ph} V_m, \qquad (10)$$

where $V_m$ is the density (mass per unit volume) of the material. The mean free path $\ell_{ph}$ of a phonon depends on (i) the Umklapp collisions with other phonons, giving to $\ell_{ph}$ a variation which is proportional to $e^{(-\frac{1}{2}\theta_D/T)}$ at low temperature tending to a constant as $T \ll \theta_D$, and (ii) the geometrical defaults and the crystal size limiting the mean free path to a length constant. Thus, for $T \ll \theta_D$, $k_{ph}$ is proportional to $c_{ph}$ and is hence a cubic function of $T$ (see Section 2.1.1).

### 2.2.2 Electronic contribution to the thermal conductivity

The electron contribution to the thermal conductivity is given by

$$k_e = \frac{1}{3} c_e v_F \ell_e V_m = \frac{1}{3} \frac{\pi^2 n k_B T}{m_e} \tau V_m, \qquad (11)$$

where $v_F$ is the Fermi velocity ($\frac{1}{2} \cdot m_e v_F^2 = E_F$), $\ell_e = v_F \tau$, the mean free path of the electrons, $\tau$ is the mean collision time, and $m_e$ is the mass of an electron. At low temperature, $\tau$ is constant and thus $k_e$ is proportional to the temperature.

### 2.2.3 Thermal conductivity of materials

The thermal conductivities of different materials are plotted in Fig. 5.

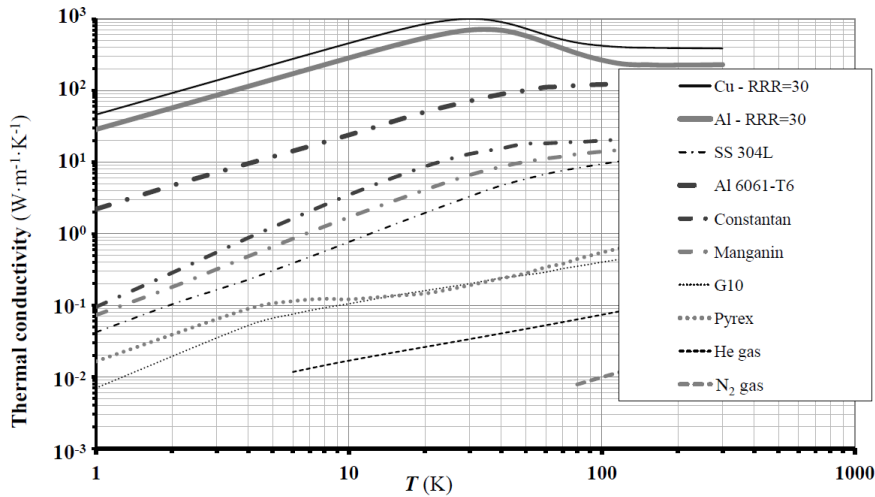

**Fig. 5:** Thermal conductivity $k$ of various materials compiled from Ref. [5]. Metals are plotted with solid lines, metallic alloys with dot-dashed lines, and thermal insulators with dotted lines. Two gases are plotted with dashed lines.

*2.2.3.1 Thermal conductivity of metals*

For pure metals, the electronic contribution dominates for any temperature: $k_e \gg k_{ph}$. At very low temperatures, $k$ is proportional to temperature. As $T$ increases, the thermal conductivity is affected by impurities included within the material. Thus, as a function of temperature, $k$ has a maximum depending on the purity of the metal: the higher the purity of the metal, the larger this maximum is, and the lower the temperature of this maximum. Figure 6 plots the thermal conductivities of coppers with different purities. The purity of the coppers is indicated by the Residual Resistive Ratio (RRR, see Section 3.1): for ordinary coppers, 5 < RRR < 150; for Oxygen-Free High thermal Conductivity (OFHC) coppers, 100 < RRR < 200; for very pure coppers, 200 < RRR < 5000.

For non-pure metals or metallic alloys, the phonon contribution may be comparable with the electronic contribution: the thermal conductivity is thus a monotonous function of temperature that is roughly linear at low temperature.

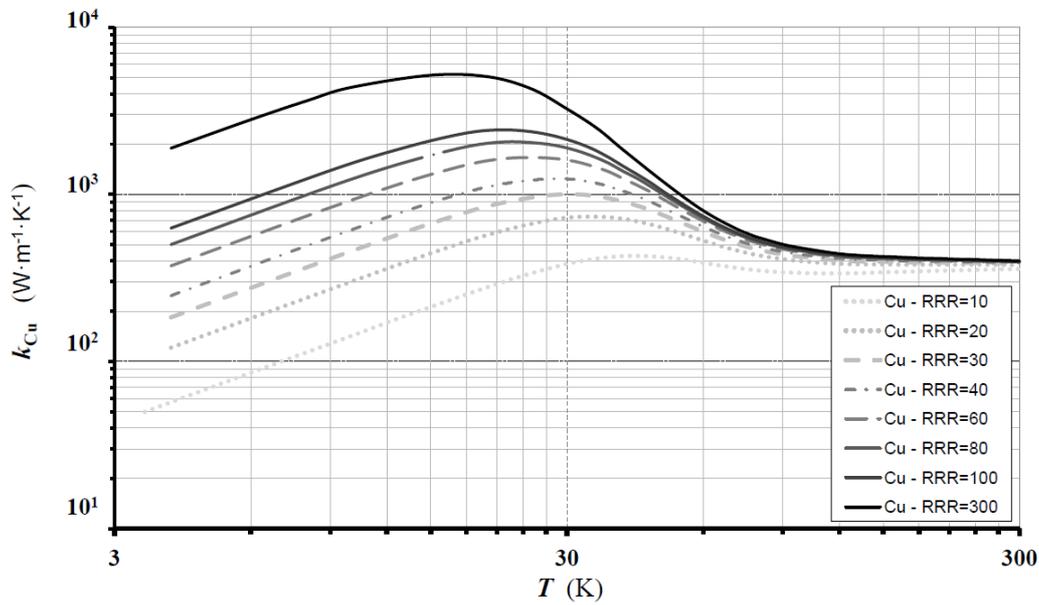

**Fig. 6:** Thermal conductivity $k_{Cu}$ of coppers having different purities, indicated by the RRR. Data were compiled from Ref. [5]. The thermal conductivities of purer coppers are plotted with darker solid lines, whereas those of less pure coppers are plotted with lighter grey dotted lines.

*2.2.3.2 Thermal conductivity of thermal insulators*

For crystallized thermal insulators, the electronic contribution is much smaller than that of the lattice contribution. Thus, their thermal conductivity is a function of $T^3$.

*2.2.3.3 Thermal conductivity of superconductors*

If $T > T_C$ (normal state), the thermal conductivity $k_s$ of superconductors behaves as for metals. If $T < T_C$ (Meissner state), $k_s$ is a function of $T^3$ and is much smaller than that of the normal state. This abrupt change in thermal conductivity is used to develop thermal interrupters.

### 2.2.4 *Integrals of thermal conductivities*

As explained in the preceding sections, because the thermal conductivity depends on temperature, especially in the cryogenic domain, one must integrate it with respect to the temperature in order to evaluate the diffusive heat flux between two temperatures (see Eq. (9)): $\int_{T_h}^{T_j} k(T)\mathrm{d}T$. Hence, for engineering purposes, it is convenient to provide integrals of the thermal conductivities evaluated from a reference temperature $T_{\text{REF}} = T_i$. They are tabulated for different materials considering $T_{\text{REF}} = 1$ K in Table A.3 of Appendix A. The integral $\int_{T_i}^{T_j} k(T) \mathrm{d}T$ can thus be evaluated by subtraction:

$$\int_{T_i}^{T_j} k(T) \mathrm{d}T = \int_{T_{\text{REF}}}^{T_j} k(T) \mathrm{d}T - \int_{T_{\text{REF}}}^{T_i} k(T) \mathrm{d}T .$$

### 2.2.5 Thermal diffusivity

Assuming no internal production of heat within a material, the heat diffusion equation is given by

$$V_{\text{m}}(T) c_p(T) \frac{\partial T}{\partial t} = \vec{\nabla} \cdot (\bar{\bar{k}}(T) \vec{\nabla} T) . \tag{12}$$

In the case of homogeneous materials, the thermal conductivity $\bar{\bar{k}}$ reduces to the scalar $k$, and if it does not depend on temperature, nor on the density $V_{\text{m}}$ and the specific heat capacity $c_p$, Eq. (12) simplifies as follows:

$$\frac{\partial T}{\partial t} = \frac{k}{V_{\text{m}} c_p} \Delta T = \kappa \Delta T , \tag{13}$$

where $\kappa$ (unit: m$^2 \cdot$s$^{-1}$) is the thermal diffusivity of the material. It can be seen as the time rate at which the (curved) initial temperature distribution is irreversibly smoothed within the material: the larger the thermal diffusivity, the faster the heat diffuses. It thus allows one to assess the time constant of the heat diffusion process. In the cryogenic domain, as thermophysical properties of a material change with temperature, the thermal diffusivity depends on the temperature, i.e. it generally increases at low temperatures; for metals, the thermal conductivity and the heat capacity are, respectively, linear and cubic functions of $T$, while the density is almost constant. The thermal diffusivities of some materials are plotted in Fig. 7. Note that the heat diffusivity of stainless steel 304L is two orders of magnitude lower than of G10 at low temperature. It follows that heat will diffuse more slowly in 304L than in G10 at low temperature; also, 304L will be conductively cooled more slowly than G10 at low temperature.

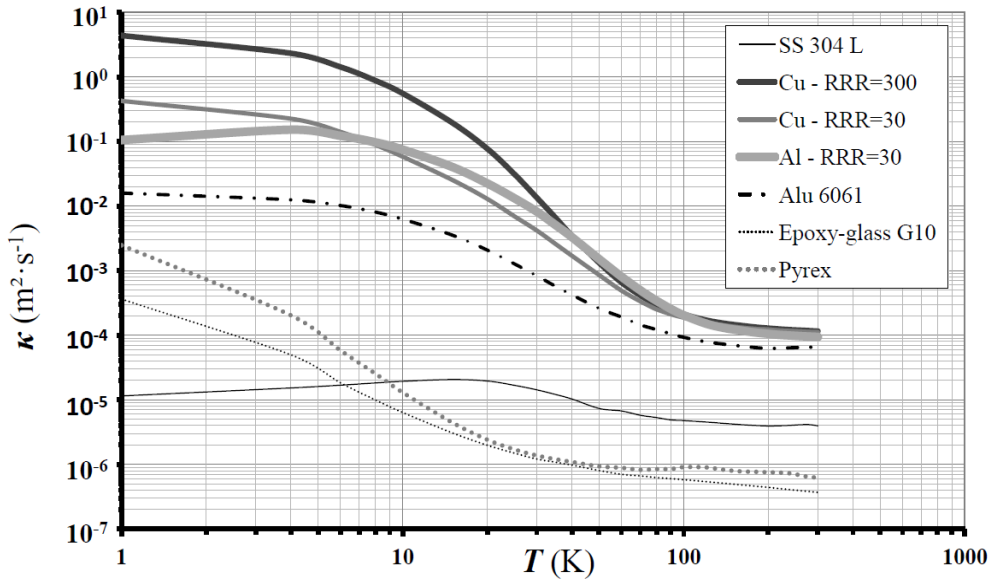

**Fig. 7:** Thermal diffusivity $\kappa$ (m$^2 \cdot$s$^{-1}$) of various materials versus the temperature $T$. Metals are plotted with solid lines and thermal insulators with dot-dashed lines. Data compiled from Ref. [5].

### 2.2.6 Thermal expansion/contraction of solids

The coefficient of expansion is defined as

$$\alpha_V = \frac{1}{V}\left(\frac{dV}{dT}\right)_P \quad \text{(unit: K}^{-1}\text{)}.$$

Being a function of the temperature, the thermal expansion can affect the resistance of an assembly, by generating large stresses, and/or the mechanical stability of an assembly (buckling). $\alpha_V$ is generally positive and so, at constant pressure, a temperature decrease induces a reduction of the physical dimensions (size) of a body. Moreover, for a solid body the effect of pressure can be neglected. The linear coefficient expansion is defined by

$$\alpha = \frac{1}{L}\frac{dL}{dT} \quad \text{(unit: K}^{-1}\text{)}. \tag{14}$$

For a crystallized solid, $\alpha$ varies as $c_{ph}$: at very low temperature it is a cubic function of the temperature and tends to a constant value as $T$ increases. In practice, the integral of the thermal expansion coefficient is computed from a reference temperature $T_{REF}$:

$$\int_{T_{REF}}^{T} \alpha(T) = \frac{\Delta L}{L} = \frac{L - L_{REF}}{L_{REF}} \quad \text{(no unit)}, \tag{15}$$

where $L_{REF}$ is the length of the body at the reference temperature. Thus, $\Delta L/L$ is proportional to $T$ around the ambient temperature and becomes proportional to $T^4$ at low temperature (in practice the coefficient of proportionality at low temperature is negligible). $\int_{T_{REF}}^{T} \alpha(T)dT$ is plotted for different materials in Fig. 8, with $L_{REF}$ at $T_{REF} = 300$ K. It can be seen that most of the thermal contraction is achieved above 50 K or 77 K (boiling point of nitrogen at a pressure of 1 atm).

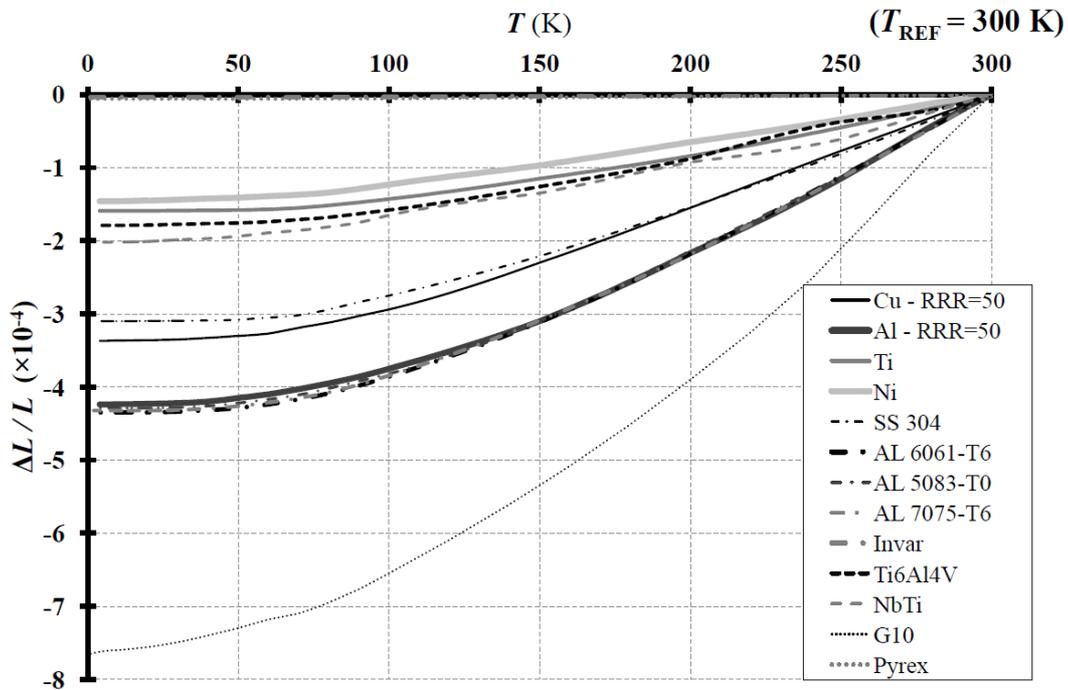

**Fig. 8:** $\Delta L/L = \int_{T_{REF}=1\,K}^{T} \alpha(T)dT$ as a function of the temperature for different solid materials

In cryogenic systems, components can be submitted to large temperature differences: at steady state because they are linked to both cold and warm surfaces (as cold mass supports, for example), or during cool-down or warm-up transient states. Special care must then be taken with assemblies of different solid materials experiencing such temperature changes in order to limit stresses and/or their destruction, as illustrated in Fig. 9. The choice of materials might determine the outcome, and/or flexibility in the assembly might be achieved by involving dedicated elements (compensation bellows, elastic or ductile elements).

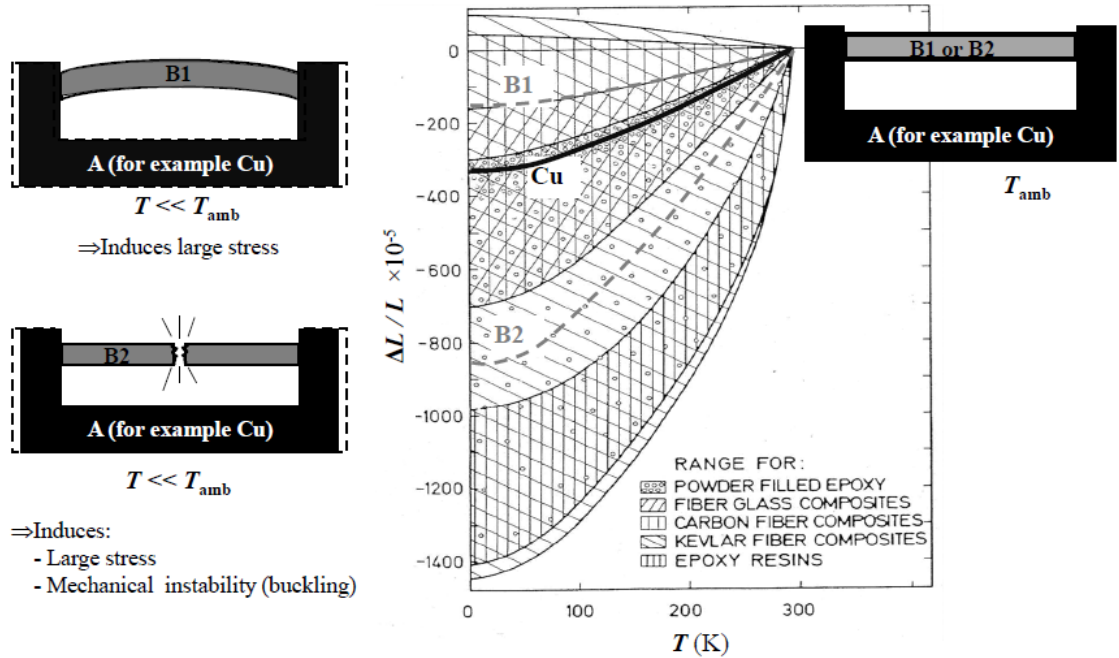

**Fig. 9:** Mechanical consequences of cooling a 'rigid' assembly of two solid materials having different thermal expansion coefficients.

## 3 Electrical conductivity/resistivity

### 3.1 Electrical resistivity of metals

For metallic conductors, electric charges $e$ are transported by $n$ 'free electrons', and the electrical conductivity is $\sigma = n \ell_e e^2 / (m_e v_F)$, where $m_e$ is the mass of an electron and $\ell_e$ denotes the mean free path of electrons. The electrical resistivity is $\rho = \sigma^{-1}$. At ambient temperature, the electron free path $\ell_e$ is dominated by electron scattering from thermal vibrations of the crystal lattice (phonons). At low temperature, it is limited mainly by scattering off chemical and physical crystal lattice imperfections (chemical impurities, vacancies, dislocations). Considering the two contributions, the electrical resistivity can be written as (Matthiesen's rule): $\rho(T) = \rho_{\text{ph}}(T) + \rho_i$, the subscript i denoting imperfections. $\rho_i$ is generally not dependent on the temperature but varies from one sample of material to another due to the imperfection content. However, $\rho_{\text{ph}}$ is the same for different samples of a material as it depends on the frequency of the collisions of electrons with thermal phonons (and electrons). For $T > \theta_D$, the concentration of phonons is proportional to the temperature, and so $\rho_{\text{ph}} \propto T$ and thus $\rho \propto T$. As $T \to 0$, $\rho \to \rho_i$, with $\rho_i$ proportional to $T^n$ (with $1 < n < 5$; see Refs. [1, 6, 7]). The electrical resistivities of several materials are plotted in Fig. 10.

As $\rho_{T \to 0}(T) = \rho_i(0)$, an indication of the impurities and crystallographic defects in a metal (imperfections) is provided by the determination of the Residual (electrical) Resistivity Ratio (RRR):

$$\text{RRR} = \frac{\rho(273 \text{ K})}{\rho(0 \text{ K})} \approx \frac{\rho(273 \text{ K})}{\rho(4.2 \text{ K})} \ . \tag{16}$$

Therefore, the smaller the imperfection content in a material, the smaller the $\rho(0) \approx \rho_i(0)$ and the larger the RRR (see the example for copper in Fig. 11).

The empirical Wiedemann–Franz law indicates that the ratio $k_e/\sigma$ has almost the same value for different metals at the same temperature. Theoretically,

$$\frac{k_e}{\sigma T} = \frac{\pi^2}{3}\left(\frac{k_B}{e}\right)^2 = 2.445 \cdot 10^{-8} \text{ (W·}\Omega\text{·K}^{-2}), \tag{17}$$

and thus, the more pure a material is, the larger its electronic thermal conductivity (see Section 2.2.3.1).

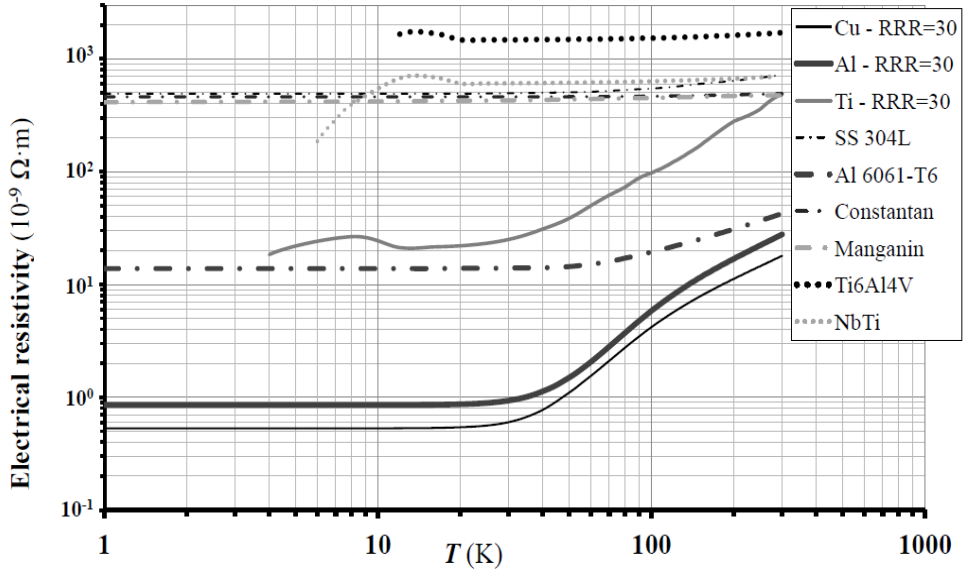

**Fig. 10:** Electrical resistivity $\rho$ ($\Omega$·m) of various materials versus temperature $T$. Metals are plotted with solid lines, metallic alloys with dot-dashed lines, and two superconducting alloys with dotted lines.

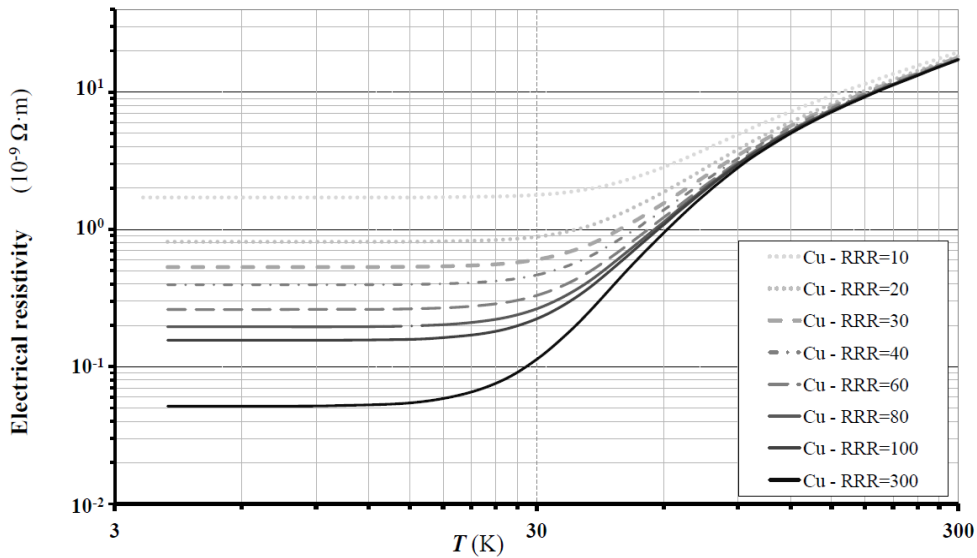

**Fig. 11:** Electrical resistivity $\rho$ of different coppers [5]. Impurities and crystallographic defect content are indicated by the RRR. Electrical resistivities of purer coppers are indicated by darker solid lines, whereas those of less pure coppers are plotted with lighter grey dotted lines.

## 3.2 Electrical resistivity of semiconductors

For semiconductors, electric charges are transported by conduction band electrons and holes in the valence band. Around the ambient temperature, lattice vibrations are predominant and electrical properties are not modified by impurities. The electrical resistivity can be expressed as $\rho(T) = a \cdot e^{\delta/2k_B T}$, where $a$ is an experimental constant and $\delta$ is the energy band depending on the material [6]. At low temperature, lattice vibrations are negligible and impurities play an important role in the transport of charges. Thus, the resistivity of semiconductors is very non-linear: it typically increases as temperature drops due to there being fewer electrons in the conduction band. An important application of this property is the use of semiconductors as temperature sensors (thermistors). Figure 12 presents the electrical resistance of some metallic and semiconductor thermistors.

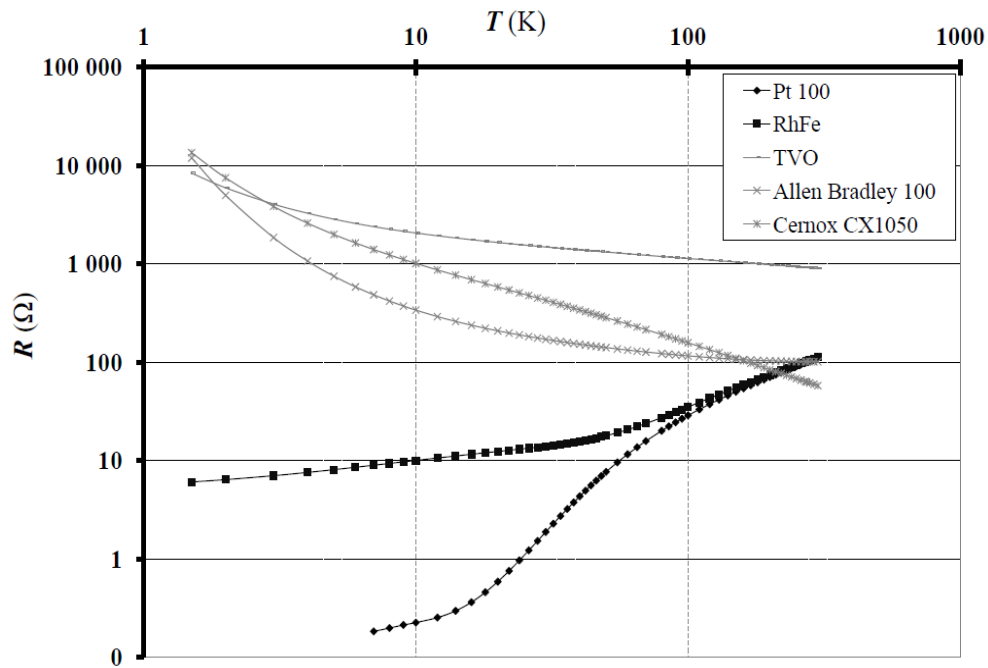

**Fig. 12:** Resistivity of various thermistors (from [8]). Resistivities of metallic sensors (Pt, RhFe) are plotted as solid black symbols; resistivities of semiconductor sensors (TVO, Allen Bradley, Cernox) are plotted as grey symbols.

## 4 Mechanical properties

In solids, atoms are in equilibrium positions depending on their interaction energies. Moving them from their equilibrium positions requires imposing an external force to overcome the restoring force. For small displacement amplitudes, the restoring force can be considered as proportional to this elastic deformation, and the equilibrium position is recovered if the external force is removed. For a larger displacement (larger external force), a limited number of atomic bonds break. Those dislocations enable atoms in specific crystal planes to slip one from another, yielding a permanent plastic deformation that propagates within the material along the densest planes of atoms because the (shear) stress needed to move increases with the spacing between the planes.

Useful macroscopic mechanical properties can be obtained by performing a tensile test (see Fig. 13) during which a homogeneous standardized test specimen is subjected to a controlled tension force $F$, increased monotonically, and then stretched until it breaks. Considering the initial dimensions of the specimen (cross-section $S_0$ and length $L_0$), the average linear stress $s = F/S_0$ and strain $\varepsilon = \Delta L / L_0 = (L - L_0)/L_0$ are plotted to outline the engineering stress–strain curve. True stress–strain curves can also be plotted considering the volume conservation of the specimen, $S_0 L_0$, and measuring the real cross-section $S$, which reduces during necking.

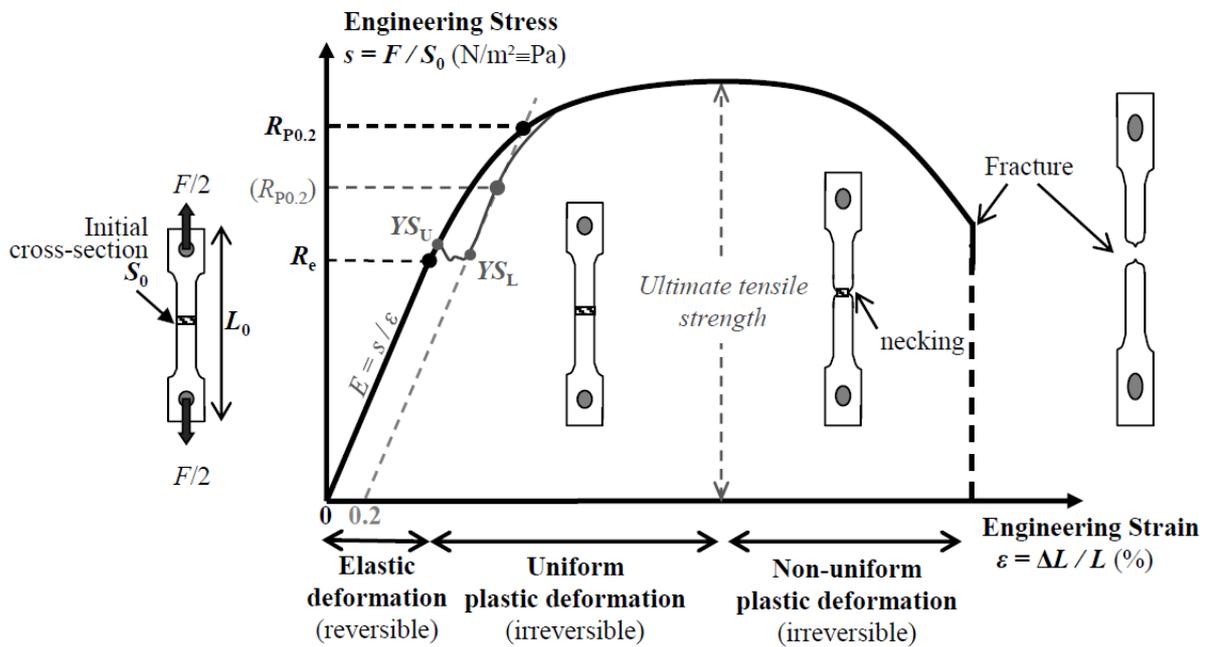

Fig. 13: Tensile test of a specimen: engineering stress–strain curve

## 4.1 Elastic domain

In the elastic domain, the elongation is proportional to the tensile force (Hooke's law), and the specimen can reversibly recover its initial size by removing the force. The proportional coefficient is the Young modulus (modulus of elasticity) defined as $E = s/\varepsilon$ (Pa); it is a measure of the stiffness $E\, S_0 / L_0$ of the material, and is used in computing structural deflections. The modulus of elasticity is determined by the binding forces between atoms and is thus generally not affected by impurities within the lattice crystal (chemical additions to obtain alloys, for example). However, temperature does affect these forces due to the thermal excitation of the lattice: the modulus of elasticity is slightly larger at low temperature than around ambient and drops at high temperature (see Fig. 14).

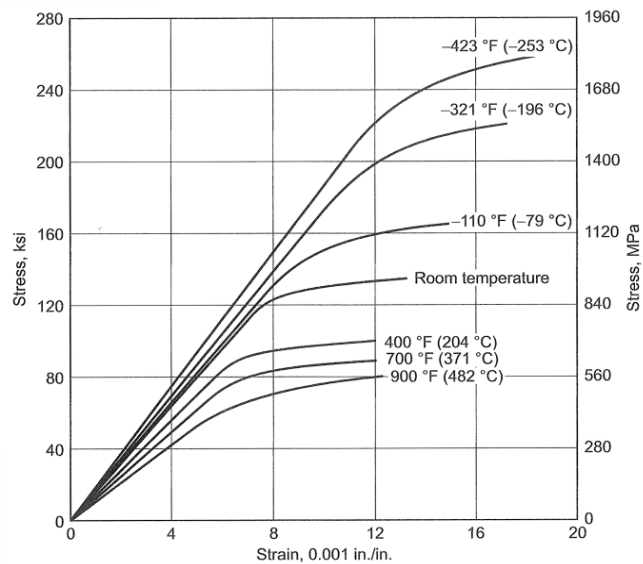

Fig. 14: Tensile stress–strain curve of annealed Ti-6Al-4V at different temperatures (from Ref. [9])

The elastic limit or yield stress, $R_e$, is the tensile stress from which elongation of the sample begins to increase disproportionately with increasing tensile force. A plastic elongation remains after the removal of the force. The yield point is not always a sharply defined point, especially for hard materials, as stress concentrations alternate with local line or band deformation steps within the material. The elastic to plastic transition can exhibit drops and fluctuations on the engineering stress–strain curve (see the grey solid line in Fig. 13). It leads to the definition of the upper- and lower-limit yield stresses $YS_U$ and $YS_L$. For engineering purposes, offset yield strengths are often used: they are the tensile stresses from which a permanent plastic deformation of 0.1%, 0.2% or 0.5% remains, and are respectively referred to as proof strengths $R_{P0.1}$, $R_{P0.2}$, and $R_{P0.5}$. The yield strength generally increases at low temperature (see Fig. 15).

Resilience is the ability of a material to absorb energy when deformed elastically and to return it when unloaded. The maximum resilience occurs at the yield point, and is called the 'resilience modulus', $U_R = \frac{1}{2} R_e^2 / E$ (J·m$^{-3}$). It is the area under the linear part of the engineering stress–strain curve. A resilient material (used to make a spring, for example) thus has a large yield stress and a small modulus of elasticity. As the temperature decreases (the relative variation of $R_e$ being generally greater than that of the modulus of elasticity (see Fig. 15)), the resilience modulus increases at low temperature.

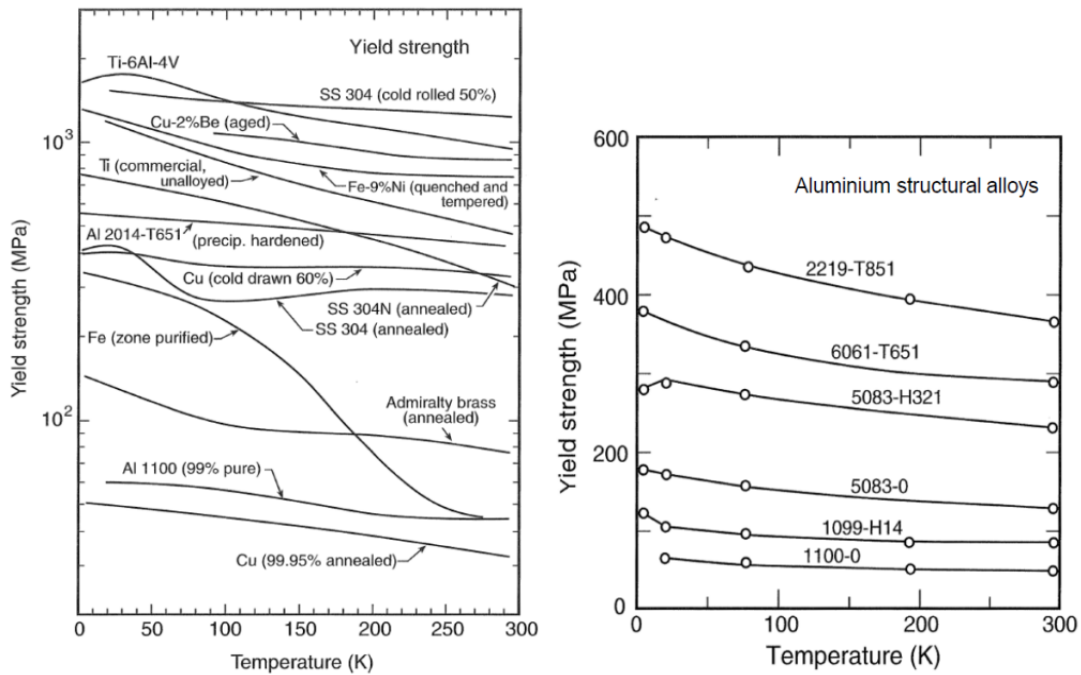

**Fig. 15:** Yield strength of various materials changing with temperature (from Ref. [9])

## 4.2 Plastic domain

The ultimate strength is the maximum tensile stress a material can withstand (under uniaxial loading), but it is not often used in practice. A material is said to be ductile when it can be plastically deformed without breaking. Otherwise it is said to be brittle. Ductility allows a material to be submitted to metal working without breaking, and gives the engineer a security margin when excess loading is applied to the material. It can be assessed by measuring the elongation and cross-section at fracture. The ductility of metals and metallic alloys depends on their crystal structure and temperature (see Figs. 16 and 17). Materials with a face-centred cubic (f.c.c.) crystal structure (such as Cu-Ni alloys, aluminium and its alloys, austenitic stainless steel, Ag, Pb, brass, Au, Pt, inconel), are ductile even at low temperatures. Thus, they are favoured in cryogenics applications, as ductility provides some safety margin from rupture. For materials with a body-centred cubic (b.c.c.) crystal structure (ferritic steels, carbon steel, steel with Ni < 10%, Mo, Nb, Cr, Nb–Ti), a ductile–brittle transition appears at low temperature: the plasticity capacity is wiped out. For other structures (Zn, Be, Zr, Mg, Co, Ti alloys etc.), no general trend appears: mechanical properties depend on interstitial components and on the symmetry axis.

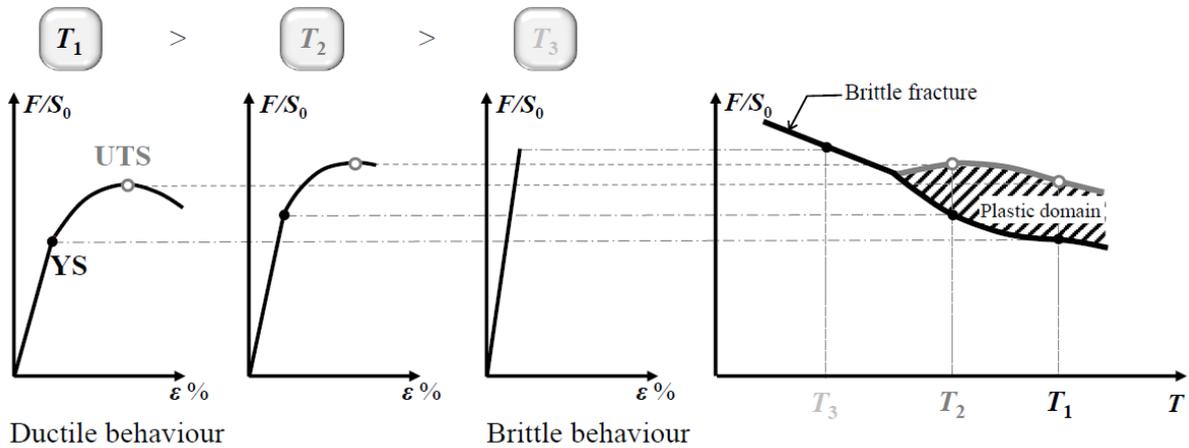

**Fig. 16:** Ductile to brittle transition of some b.c.c. atomic lattice materials when cooled down

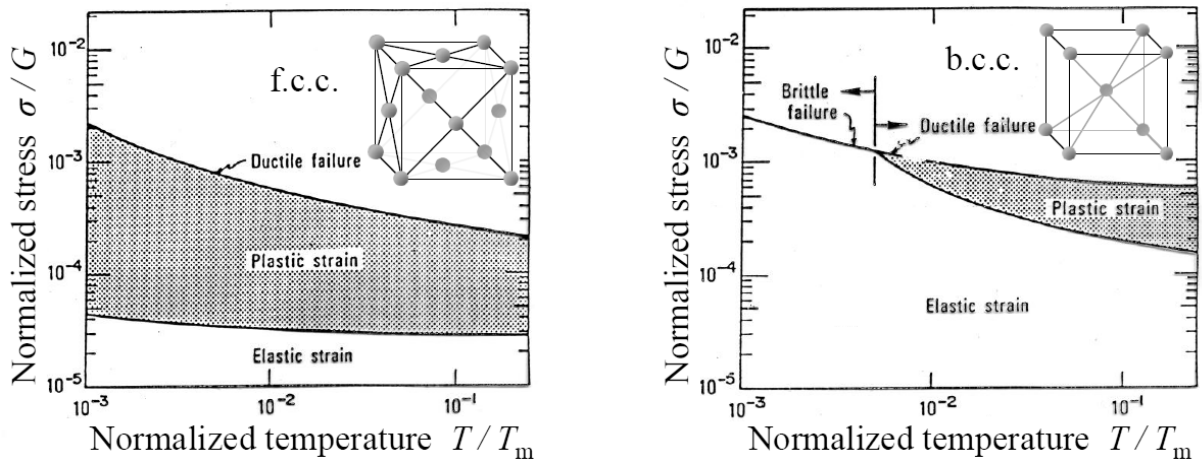

**Fig. 17:** General trend of yields for f.c.c. and b.c.c. materials. Temperature is normalized by the melting temperature $T_m$; stress is normalized by the shear modulus $G$. Originally from Ref. [10].

The toughness of a material describes its ability to absorb energy and plastically deform until fracture. A tough material can withstand occasional stresses above the yield stress without fracturing. The toughness of f.c.c. materials can be judged from tensile tests [11]. The modulus of toughness UT is the amount of work per unit volume (J·m$^{-3}$) that can be given to a material without causing it to fracture, and is the total area under the stress–strain curve. Toughness is thus based on strength and ductility. Notch or fracture toughness is the ability of a notch specimen to absorb elastic or kinetic energy up to fracture. A tensile test or a high-strain-rate test, such as the Charpy impact test, are used to measure it. It is expressed by way of a critical stress intensity factor $K$ (Pa·m$^{0.5}$), at which point a thin crack begins to grow. It is the resistance of a material to brittle fracture when a crack exists. As the temperature decreases, both toughness and fracture toughness decrease. As shown in Fig. 18, the values of fracture toughness of high-strength nickel and titanium b.c.c. alloys drop at low temperature, whereas the value remains large for f.c.c. stainless steel 310.

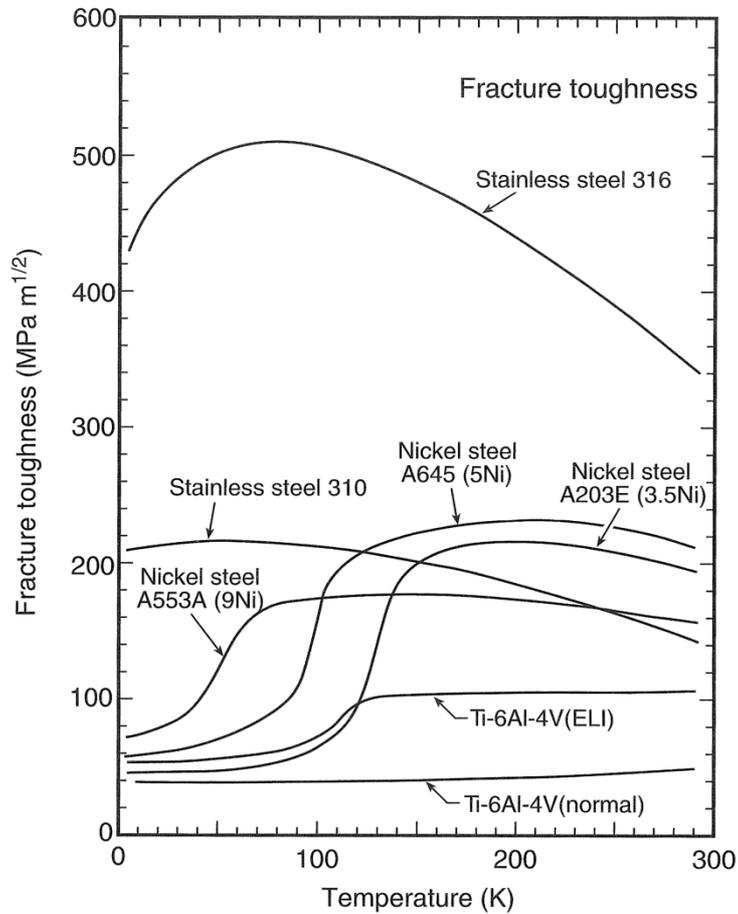

**Fig. 18:** Fracture toughness of materials: plane-strain critical-stress intensity factor for cracking mode I (named "opening mode" and for which a tensile stress normal to the plane of the crack is applied). From Ref. [9].

## 5  Magnetic properties

In a vacuum, the magnetic field $\vec{B}$ (unit: $T \equiv V \cdot s \cdot m^{-2} \equiv N \cdot A^{-1} \cdot m^{-1}$) is proportional to the excitation magnetic field $\vec{H}$ (unit: $V \cdot s \cdot A^{-1} \cdot m^{-1} \equiv A \cdot m^{-1}$): $\vec{B} = \mu_0 \vec{H}$, the proportional factor being the permeability of free space $\mu_0 = 4\pi \cdot 10^{-7}\,(N \cdot A^{-2})$. In a material, $\vec{B} = \mu_0(\vec{H} + \vec{M})$, where $\vec{M} = \chi \vec{H}$ (unit: $A \cdot m^{-1}$) is the magnetization and represents how strongly a region of material is magnetized by the application of $\vec{H}$. It is the net magnetic dipole moment per unit volume. Thus, $B = \mu_0(1+\chi)H = \mu_0 \mu_r H$, where $\mu_r$ is the relative magnetic permeability. The magnetic moment of a free atom depends on electron spin, the orbital kinetic moment of the electrons around the nucleus, and the kinetic moment change induced by the application of a magnetic field. Five types of magnetic behaviour can be distinguished: diamagnetism and paramagnetism, which are induced by isolated atoms (ions) and free electrons, and ferromagnetism, antiferromagnetism, and ferrimagnetism, which occur due to the collective behaviour of atoms.

### 5.1  Diamagnetic materials

If the magnetic susceptibility $\chi = \mu_r - 1 < 0$, it causes a diamagnet to create a magnetic field in opposition to an externally applied magnetic field. When the field is removed, the effect disappears. Examples of diamagnetic materials are silver, mercury, diamond, lead, and copper. If the (small) field $\vec{H}$ is applied, then $\vec{M} = \chi \vec{H}$ and $\chi$ does not depend on temperature. Type I superconductors (such as copper and niobium) are perfect diamagnets for $T < T_C$.

## 5.2 Paramagnetic materials

A magnetic susceptibility $\chi = \mu_r - 1 > 0$ (with $\chi$ small) induces paramagnets (such as aluminium) to be slightly attracted by an externally applied magnetic field. Since the spin of unpaired electrons in the atomic orbitals is randomized, the induced permanent magnetic moment (dipoles) has no effect. However, if a magnetic field is applied, the dipoles tend to align with the applied field, yielding a net magnetic moment. When the field is removed, the effect disappears. For low levels of magnetization, $\chi = \mathcal{C}/T$, where $\mathcal{C}$ is the Curie constant. An application of this dependence of the magnetic susceptibility with temperature is the principle of magnetic thermometers.

## 5.3 Ferromagnetic materials

For ferromagnetic materials, such as Fe-Ni-Co alloys (not austenitic steels), magnetization is due to the contribution of the unpaired electron spins (as for paramagnets) and coupling interactions causing magnetic moments of adjacent atoms to be parallel to an applied magnetic field and parallel to each other. Thus, magnetization remains.

Above the Curie temperature $\theta_\mathcal{C}$, which is generally larger than the ambient temperature, the Curie–Weiss law yields a paramagnetic behaviour: $\chi = cst/(T - \theta_\mathcal{C})$.

Below $\theta_\mathcal{C}$, $(M(0) - M(T))/M(0) \propto T^{3/2}$. When an increase in the applied external magnetic field $H$ cannot augment the magnetization $M$, the material reaches its saturation state. Figure 19 shows the saturation magnetization $M_s$ of nickel as a function of (normalized) temperature.

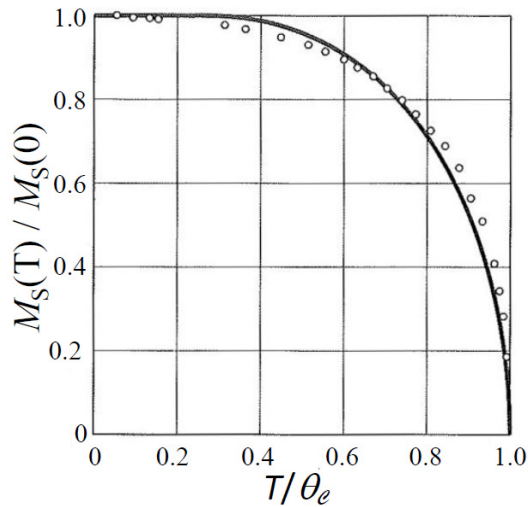

**Fig. 19:** Magnetization at saturation of nickel as a function of temperature (from Ref. [1])

## 5.4 Antiferromagnetic materials

For antiferromagnets, the tendency of the magnetic moments of neighbouring atoms is to point in opposite directions. A substance is antiferromagnetic when all atoms are arranged such that each neighbour is 'anti-aligned'. Antiferromagnets have a zero net magnetic moment below a critical temperature called the Néel temperature, $\theta_N$, and thus $B = 0$ for $T < \theta_N$. Antiferromagnets can exhibit diamagnetic and ferrimagnetic properties for $T > \theta_N$: $\chi = cst/(T + \theta_N)$.

## Appendix A: Tabulated quantities

**Table A.1:** Debye temperatures of various metals

| Material | $\theta_D$ (K) |
|---|---|
| Copper | 343 |
| Aluminium | 428 |
| Titanium | 428 |
| Niobium | 275 |
| Nickel | 450 |
| AISI304 | 470 |
| AISI316 | 500 |

**Table A.2:** $\gamma$ coefficient of the linear dependency of the heat capacity of various metals on temperature

| Material | $\gamma$ ($10^{-3}$ J·kg$^{-1}$·K$^{-2}$) |
|---|---|
| Copper | 11.0 |
| Aluminium | 50.4 |
| Titanium | 74.2 |
| Niobium | 94.9 |
| Nickel | 124.0 |

Table A.3: $\int_{T_i}^{T_j} k(T) dT$ of different materials; data compiled from [5] and considering $T_{REF} = 1$ K

| T (K) | SS304 | Cu-RRR=300 | Cu-RRR=30 | Brass | Constantan | Manganin | Invar | Ti-6Al-4V | Al-RRR=30 | 6061-T6 | 5083-T0 | Nb–Ti |
|---|---|---|---|---|---|---|---|---|---|---|---|---|
| 2 | 0.073 | | 69 | 1.0 | 0.18 | 0.124 | 0.0388 | 0.17 | 42.8 | 3.5 | 1.06 | 0.04 |
| 4 | 0.40 | 3560 | 345 | 6.0 | 1.31 | 0.773 | 0.276 | 0.80 | 214 | 17.7 | 5.61 | 0.27 |
| 6 | 1.02 | 8360 | 807 | 16.0 | 3.87 | 2.12 | 0.819 | 1.78 | 501 | 41.4 | 13.7 | 0.76 |
| 8 | 1.96 | 14 900 | 1450 | 31.0 | 8.03 | 4.22 | 1.7 | 3.07 | 900 | 74.6 | 25.3 | 1.5 |
| 10 | 3.3 | 22 800 | 2270 | 51 | 13.9 | 7.1 | 2.95 | 4.67 | 1410 | 118 | 40.5 | 2.5 |
| 15 | 8.5 | 46 600 | 5130 | 128 | 38.2 | 18.6 | 7.93 | 9.91 | 3190 | 272 | 95.2 | 6.0 |
| 20 | 16.7 | 72 900 | 8910 | 235 | 74.8 | 35.9 | 15.6 | 16.7 | 5590 | 487 | 173 | 11.1 |
| 25 | 28.1 | 95 800 | 13 500 | 370 | 124 | 59.7 | 26 | 24.7 | 8560 | 765 | 273 | 17.4 |
| 30 | 42.8 | 115 000 | 18 400 | 525 | 184 | 89.6 | 39.1 | 33.8 | 11 900 | 1100 | 395 | 25.2 |
| 35 | 61.2 | 130 000 | 23 300 | 697 | 252 | 125 | 54.7 | 43.8 | 15 400 | 1480 | 538 | 34.4 |
| 40 | 82.9 | 140 000 | 28 000 | 883 | 328 | 166 | 72.9 | 54.8 | 18 900 | 1900 | 701 | 45.0 |
| 50 | 136 | 155 000 | 36 200 | 1280 | 497 | 260 | 117 | 79.4 | 25 300 | 2840 | 1080 | 70.4 |
| 60 | 199 | 164 000 | 42 900 | 1730 | 679 | 367 | 170 | 107 | 30 500 | 3900 | 1540 | 101 |
| 70 | 271 | 171 000 | 48 400 | 2210 | 865 | 483 | 232 | 137 | 34 800 | 5020 | 2050 | 138 |
| 77 | 326 | 176 000 | 51 800 | 2580 | 997 | 569 | 281 | 160 | 37 300 | 5830 | 2440 | 167 |
| 80 | 350 | 177 000 | 53 300 | 2740 | 1050 | 607 | 302 | 171 | 38 300 | 6180 | 2610 | 180 |
| 90 | 436 | 182 000 | 57 800 | 3320 | 1250 | 739 | 379 | 207 | 41 400 | 7370 | 3220 | 228 |
| 100 | 527 | 187 000 | 62 000 | 3950 | 1440 | 877 | 462 | 245 | 44 200 | 8580 | 3870 | 280 |
| 120 | 725 | 196 200 | 70 270 | 5330 | 1847 | 1165 | 640 | 329 | 49 240 | 11 040 | 5280 | 398 |
| 140 | 940 | 204 900 | 78 200 | 6860 | 2269 | 1467 | 834 | 422 | 53 830 | 13 560 | 6820 | 530 |
| 160 | 1170 | 213 300 | 86 100 | 8500 | 2700 | 1781 | 1040 | 522 | 58 300 | 16 130 | 8490 | 673 |
| 180 | 1414 | 221 700 | 94 000 | 10 240 | 3140 | 2107 | 1258 | 630 | 62 800 | 18 780 | 10 270 | 824 |
| 200 | 1667 | 229 900 | 101 800 | 12 080 | 3600 | 2447 | 1482 | 744 | 67 300 | 21 480 | 12 170 | 983 |
| 220 | 1937 | 238 200 | 109 600 | 13 950 | 4060 | 2797 | 1732 | 865 | 71 800 | 24 180 | 14 170 | 1150 |
| 240 | 2207 | 246 300 | 117 400 | 16 050 | 4530 | 3167 | 1982 | 993 | 76 400 | 27 080 | 16 370 | 1323 |
| 260 | 2487 | 254 400 | 125 100 | 18 150 | 5000 | 3557 | 2242 | 1127 | 80 900 | 30 080 | 18 570 | 1503 |
| 280 | 2777 | 262 500 | 132 900 | 20 350 | 5480 | 3967 | 2502 | 1265 | 85 400 | 33 180 | 20 970 | 1687 |
| 300 | 3077 | 270 500 | 140 600 | 22 650 | 5970 | 4397 | 2772 | 1415 | 90 000 | 36 380 | 23 470 | 1875 |